\pdfoutput=1
\documentclass{aa}
\nolinenumbers
\renewcommand{\linenumbers}{}
\usepackage{graphicx} 
\usepackage[varg]{txfonts}
\usepackage{hyperref}

\bibpunct{(}{)}{;}{a}{}{,} 

\titlerunning{Faster $\chi^2$ periodogram}
\authorrunning{K. Rutkowski \& R. Poleski}

\begin{document}

\title{Faster $\chi^2$ periodogram}
\subtitle{Accelerating linear algebra with fast Toeplitz solvers}
\author{K. Rutkowski\inst{1}\corrauth{kk.rutkowski@student.uw.edu.pl} \and R. Poleski\inst{1}\email{rpoleski@astrouw.edu.pl}}

\institute{Astronomical Observatory, University of Warsaw, Al. Ujazdowskie 4, 00-478 Warszawa, Poland}

\date{Received date /
Accepted date}

\abstract {In recent years, the traditional Lomb--Scargle periodogram, relying on least-squares fitting of a sinusoid to data with a known mean, has been mostly superseded by its floating-mean extension---the generalized Lomb--Scargle periodogram---due to the superior robustness and accuracy of the latter. While further generalization to higher-degree trigonometric polynomials (known as the multiharmonic analysis of variance and the fast $\chi^2$ periodogram) has been known for decades, its usage remains relatively uncommon, mostly due to its significantly higher computational cost. Here we present a technique reducing the computational complexity of such an extension by combining a Toeplitz formulation of the system of normal equations with non-uniform fast Fourier transforms, coupled with condition-number monitoring to protect against numerical instability. When modified for practical efficiency rather than asymptotic computational complexity, our approach reaches speed-ups exceeding an order of magnitude over current state-of-the-art implementations across a wide range of hyperparameters when evaluated on multithreaded systems, while matching or exceeding their accuracy. Additionally, we provide a reference implementation used for benchmarking, along with the code used to generate the results presented in this work.}

\keywords{methods: data analysis -- methods: statistical -- methods: numerical -- stars: oscillations}
\maketitle

\section{Introduction}
\label{sec:introduction}

The Lomb--Scargle periodogram \citep{Lomb1976, Scargle1982} is among the most widely used tools in astronomical time-series analysis, largely owing to its computational efficiency. Both formulations are mathematically equivalent up to normalization and rely on least-squares fitting of a zero-mean sinusoid to irregularly sampled data.

Although such approaches are often sufficient to detect periodic variability, they are limited by their assumption that the mean value of the data equals the average value of the fitted model. This limitation was partially resolved by floating-mean periodograms \citep[e.g.,][]{1981AJ.....86..619F, schwartzenberg-czerny1996, GLS}, whose computational cost is directly comparable to that of their zero-mean counterparts, while offering improved frequency determination and reduced susceptibility to aliasing. In particular, under observational conditions that are by no means rare, the zero-mean periodogram of \citet{Lomb1976} can lose a substantial fraction of the signal power, whereas floating-mean methods that fit a constant offset---such as the generalized Lomb--Scargle periodogram---do not suffer from this loss \citep[see][fig.~1]{1995AJ....109.1889F}.

Despite these advantages, the practical application of floating-mean periodograms remains limited by the assumption of sinusoidal variability. While multiharmonic extensions of these periodograms---the multiharmonic analysis of variance (AoVMH) and the fast $\chi^2$ periodogram\footnote{The name ``fast $\chi^2$'' refers to computing the least-squares residual $\chi^2(f)$ via an FFT-based algorithm and does not imply a $\chi^2$-distributed test statistic; the variance-ratio statistic corresponding to the coefficient of determination ($R^2$) follows Fisher's $F$-distribution \citep{Scheffe1959, schwartzenberg-czerny1996} under the assumption that the measurements follow a Gaussian distribution.}---have been known for decades and offer increased robustness and accuracy \citep{schwartzenberg-czerny1996, palmer2009}, their adoption has remained limited, partly due to their computational cost. In general, na\"{\i}ve least-squares fitting of a degree-$d$ trigonometric polynomial (equivalent to fitting a Fourier series truncated at $d$ harmonics to a light curve containing $M$ measurements) across $N$ frequency nodes requires $\mathcal{O}(dN(M + d^2))$ operations. To reduce this computational burden, two fast algorithms have been developed. The Szeg\H{o} recursion approach shown by \citet{schwartzenberg-czerny1996} allows an exact solution of such a system to be derived in $\mathcal{O}(dNM)$ operations by omitting the dense linear solve; however, because the sums are evaluated directly in the time domain, its cost still scales linearly with $M$ per trial frequency. An alternative approach computes an arbitrary-precision approximation by evaluating the trigonometric sums via a single type-1 non-uniform fast Fourier transform (NuFFT-1; \citealt{palmer2009}), allowing the algorithm to achieve a computational complexity of $\mathcal{O}(dN(\log N + d^2))$, sublinear with respect to $M$, at the cost of retaining an $\mathcal{O}(d^3)$ per-frequency-node dense linear solve. This NuFFT-based class also includes the state-of-the-art implementations against which we benchmark in Sect.~\ref{sec:comparison}, namely \textsc{nifty-ls} \citep{Garrison_2024} and the fast Astropy periodogram implementations \citep{astropy}.

\begin{figure}
  \resizebox{\hsize}{!}{\includegraphics{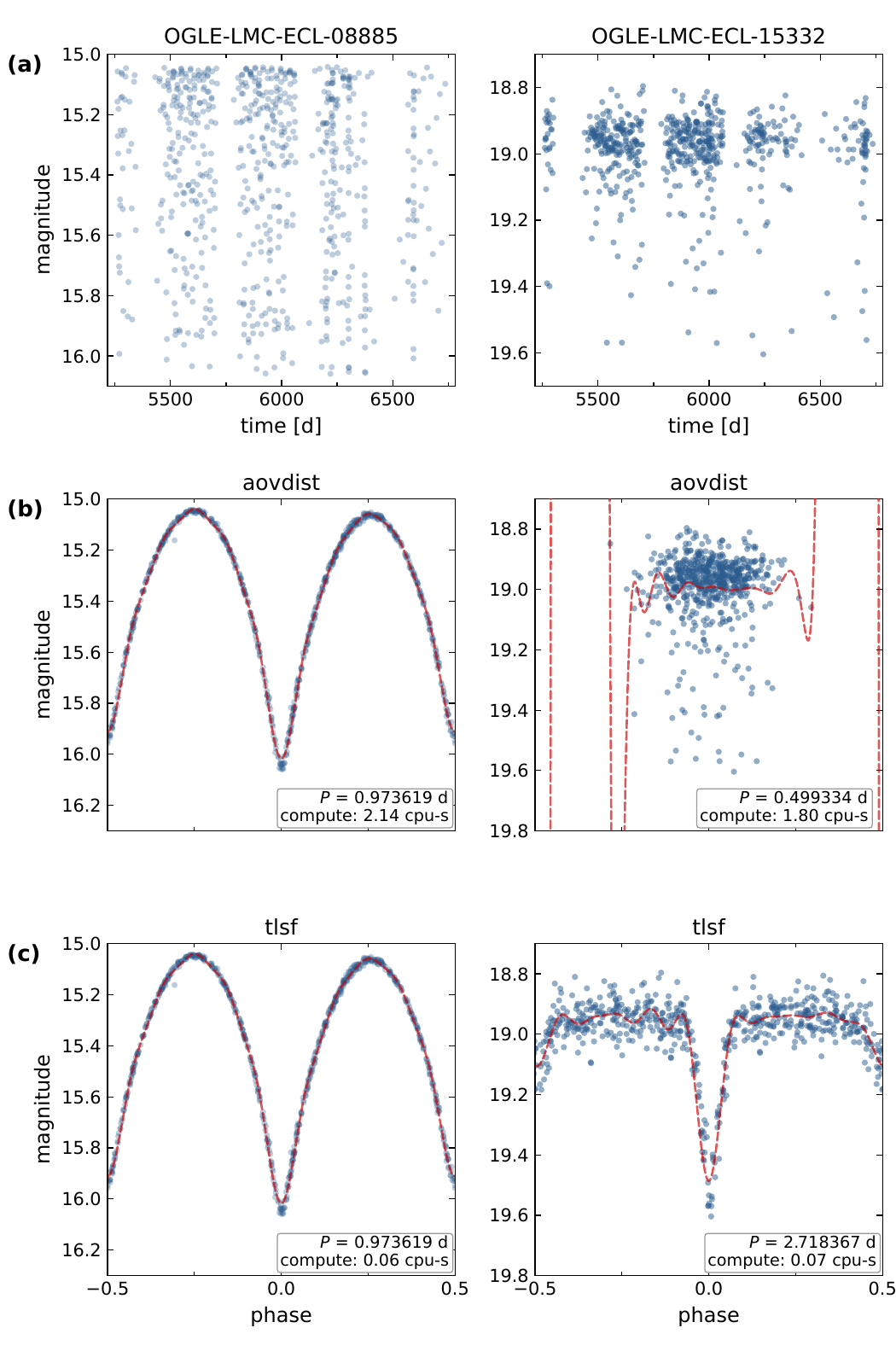}}
  \caption{Light curves of OGLE-LMC-ECL-08885 (left) and OGLE-LMC-ECL-15332 (right) with fitted $d=11$ models (dashed curves). The rows show (a) the raw light curves with observation times in $\mathrm{HJD} -2\,450\,000$, measured in days ($\mathrm{d}$), and fits folded at the peak periods $P$ from (b) \textsc{aovdist} and (c) \textsc{tlsf}, with zero phase selected at the moment of minimum brightness and single-periodogram computation times given in each phased panel. OGLE-LMC-ECL-15332 was explicitly selected because the peak period returned by \textsc{aovdist} differs from that of our implementation, and this choice does not represent the typical numerical stability of either of the implementations.}
  \label{fig:two_lightcurves}
\end{figure}

In this work, we show that expressing the trigonometric polynomial in its complex exponential form reduces the NuFFT-1-based normal equations to a Hermitian Toeplitz system. This brings the computational cost down to $\mathcal{O}(dN(\log M + \log^2 d))$ operations via superfast Toeplitz solvers, or to $\mathcal{O}(dN(\log M + d))$ via the Levinson recursion---strictly lower than both previously known fast approaches. Additionally, we discuss the effect of the fast Fourier transform (FFT) block size on empirical execution time and examine the numerical stability of this approximation under condition-number monitoring. Figure~\ref{fig:two_lightcurves} illustrates the practical impact of this approach on two representative eclipsing binary light curves from the Optical Gravitational Lensing Experiment \citep[OGLE-IV;][]{LMC-ECL}, comparing our single-precision implementation (\textsc{tlsf}) against the reference AoVMH implementation (\textsc{aovdist}; \citealt{schwartzenberg-czerny1996}): both solvers recover the catalog period of OGLE-LMC-ECL-08885, whereas condition-number monitoring in \textsc{tlsf} avoids the daily alias selected by \textsc{aovdist} for OGLE-LMC-ECL-15332 at a fraction of the computational cost. Additionally, we examine the effects of matrix conditioning and the parallel scalability of this approach.

\section{Computational complexity of the non-uniform fast Fourier transform}
\label{sec:nufft_complexity}

Throughout this paper, we consider the analysis of an unevenly sampled time series consisting of $M$ observations with observation times $t_m$, measured values $y_m$, and statistical weights $w_m$, for $m=1,\dots,M$.
The frequency is denoted by $f$, with a target grid of $N$ frequencies, of which $B$ are evaluated within a single FFT block using frequency step $\Delta_f$.

Additionally, we adopt the type-1 NuFFT formulation \citep{dutt1993} as named by \citet{greengard2004}, following the application of non-uniform fast Fourier methods to least-squares periodograms pioneered by \citet{Press1989}. Under this convention, a NuFFT-1 evaluates sums corresponding to non-uniform measurement times on an equispaced frequency grid
\begin{equation}
    f_j = j\,\Delta_f, \qquad j=0,\dots,B-1.
\end{equation}
For a block shifted by $f_0$, application of the shift theorem allows us to express the complex Fourier sum $X_{f_0,j}$ evaluated at frequency $f_0 + j\,\Delta_f$ via the NuFFT-1 as
\begin{equation}
\begin{aligned}
    X_{f_0,j}
    &= \sum_{m=1}^{M} y_m e^{-2\pi i(f_0+j\,\Delta_f)t_m} \\
    &= \sum_{m=1}^{M}
    \left(y_m e^{-2\pi i f_0 t_m}\right)
    e^{-2\pi i j\,\Delta_f\,t_m}.
\end{aligned}
\end{equation}
Thus, $f_0$ is applied by complex multiplication of the input values, while the internal transform remains type-1, as the frequencies $j\,\Delta_f$ remain equispaced after the shift operation.

While many different algorithms for evaluating the NuFFT have been proposed, here we follow the convolution-based algorithm developed by \citet{dutt1993}. However, we note that the optimization procedure described below can be applied to alternative approaches, such as the low-rank approximation-based evaluation \citep{LRA_NFFT}, without requiring any modifications.

\subsection{Theoretical computational complexity}

In the asymptotic regime ($N\to\infty$ and $M\to\infty$) and neglecting constant prefactors, the number of operations required to evaluate the $N$ frequencies in blocks of size $B$ is proportional to
\begin{equation}
    \mathcal{C}_{\mathrm{op}}(B)
    = N\log B+\frac{NM}{B}+B,
\end{equation}
where the first term is the FFT operation count, the second term models the repeated frequency shifts of the $M$ measurements over $N/B$ blocks, and the final term represents the precomputation cost shared by all blocks. Differentiating with respect to $B$ gives $\partial \mathcal{C}_{\mathrm{op}}/\partial B = N/B - NM/B^2 + 1 = (N/B)(1 - M/B) + 1$. In the $N/M\to\infty$ limit, the share of the precomputation per frequency disappears, and the root converges to $B_0=M$. Consequently, the resulting computational complexity of NuFFT-1 becomes
\begin{equation}
    \mathcal{C}_{\mathrm{op}} = \mathcal{O}(N\log M).
\end{equation}

\subsection{Empirical time complexity}
\label{sec:emp_complexity}
Although the formula for optimizing the operation count is relatively straightforward, in many practical applications other factors (such as memory access patterns) tend to dominate the computational cost. We therefore provide an alternative block-size optimization framework based on the concept of empirical time complexity, as defined by \citet{goldsmith2007}. Under this model, we approximate the FFT time complexity as a power-law function of the transform length and model the wall-clock execution time as
\begin{equation}
    \mathcal{C}_{\mathrm{emp}}(B) = N B^\alpha + \beta\frac{NM}{B} + \gamma B,
    \quad
    \alpha>0,\; \beta>0,\; \gamma\ge0,
    \label{eq:empirical_complexity}
\end{equation}
where $\alpha$, $\beta$, and $\gamma$ are empirically derived scaling constants. Because the $N$ frequencies are partitioned into $N/B$ separate transforms, assuming a $B^{1+\alpha}$ runtime per transform, analogously to the power-law scaling behavior reported by \citet{goldsmith2007}, yields $(N/B)B^{1+\alpha} = N B^\alpha$ for the first term. The second and third terms retain their previous interpretations, with the scaling coefficients accounting for finite sample sizes.

Assuming $\gamma=0$ as an initial guess, differentiating with respect to $B$ yields
\begin{equation}
    \mathcal{C}_{\mathrm{emp}}'(B)
    = \alpha N B^{\alpha-1} - \beta\frac{NM}{B^2},
\end{equation}
from which the optimal block size is given by
\begin{equation}
    \alpha B^{\alpha+1} = \beta M,
    \qquad
    B_0 = \left(\frac{\beta M}{\alpha}\right)^{1/(1+\alpha)}.
\end{equation}
In finite-sample cases ($\gamma>0$), the positive setup term shifts the root of $\mathcal{C}_{\mathrm{emp}}'(B)$ to a smaller block size, so $B_{\mathrm{opt}}\le B_0$. This term is subleading in the asymptotic limit considered here and does not change the scaling. Substituting the optimal block size gives an empirical complexity of
\begin{equation}
    \mathcal{C}_{\mathrm{emp}}(B_0)
    = \mathcal{O}\!\left(NM^{\alpha/(1+\alpha)}\right).
\end{equation}

While the optimal block size for the general case ($\gamma>0$) can be determined using numerical optimization algorithms (such as Newton's method), most of the computationally efficient FFT implementations restrict the number of frequencies per block $B$ to integers with no prime factors greater than $3$. Consequently, direct evaluation of Eq.~\eqref{eq:empirical_complexity} for suitable candidates below the initial guess $B_0$ remains the most reasonable practical approach. Additionally, we expect $\mathcal{C}_{\mathrm{emp}}$ to approximate the observed scaling more closely than the operation count $\mathcal{C}_{\mathrm{op}}$ for most hardware configurations, making the semi-empirical model preferable for practical block-size selection.

\section{Solving systems of linear equations}
\label{sec:linear_systems}

Although the NuFFT is one of the key components of the multiharmonic least-squares periodogram, it alone is not sufficient to obtain the least-squares solution. Following \citet{palmer2009}, fitting the degree-$d$ trigonometric polynomial at each target frequency $f$ reduces to solving a system of exactly $2d+1$ linear equations, whose computational cost often exceeds the computation required to evaluate the NuFFTs. In this section, we therefore focus on efficient model fitting once the NuFFTs have already been evaluated.

The model used for evaluation can generally be expressed as
\begin{equation}
    \label{eq:model}
    y_m \approx A_0 + \sum_{k=1}^{d} \left[A_k \cos(2\pi k f t_m) + B_k \sin(2\pi k f t_m)\right].
\end{equation}
For conciseness, we introduce the harmonic sums
\begin{align*}
    C_k &= \sum_{m=1}^{M} w_m\cos(2\pi k f t_m), & S_k &= \sum_{m=1}^{M} w_m\sin(2\pi k f t_m), \\
    C_k^{Y} &= \sum_{m=1}^{M} w_m y_m\cos(2\pi k f t_m), & S_k^{Y} &= \sum_{m=1}^{M} w_m y_m\sin(2\pi k f t_m),
\end{align*}
along with the frequency-independent data sums
\begin{equation}
    Y = \sum_{m=1}^{M} w_m y_m, \qquad YY = \sum_{m=1}^{M} w_m y_m^2.
\end{equation}
The zero-frequency term follows the same harmonic sequence, with
\begin{align*}
    C_0 &= \sum_{m=1}^{M} w_m, & C_0^{Y} &= Y, \\
    S_0 &= 0, & S_0^{Y} &= 0.
\end{align*}
The degree-$d$ system is therefore fully determined by the sums $C_k$ and $S_k$ up to $2d$, and by $C_k^{Y}$ and $S_k^{Y}$ up to $d$.

\subsection{Real basis formulation}

By collecting the unknown coefficients into
\begin{equation}
    \boldsymbol{\theta}(f) = (A_0,\ldots,A_d,B_1,\ldots,B_d)^\mathsf{T},
\end{equation}
minimizing the weighted least-squares residual of Eq.~\eqref{eq:model} reduces to enforcing the zero-gradient condition with respect to $\boldsymbol{\theta}(f)$. This produces the system of normal equations \citep{Bjorck1996}, which equates the weighted Gram matrix of the trigonometric basis $\mathcal{G}(f)$ to the data projection vector, namely
\begin{align*}
    \mathcal{G}(f)\boldsymbol{\theta}(f) &=
    \begin{pmatrix}
        \boldsymbol{r}^c(f) \\
        \boldsymbol{r}^s(f)
    \end{pmatrix}, \\
    \mathcal{G}(f) &=
    \begin{pmatrix}
        \mathcal{G}^{cc}(f) & \mathcal{G}^{cs}(f) \\
        \mathcal{G}^{sc}(f) & \mathcal{G}^{ss}(f)
    \end{pmatrix},
\end{align*}
where the right-hand side components are
\begin{align*}
    r_p^c(f) &= C_p^{Y}, \qquad p=0,\dots,d, \\
    r_q^s(f) &= S_q^{Y}, \qquad q=1,\dots,d.
\end{align*}
For $p,q \in \{0,\dots,d\}$ in the cosine block and $p,q \in \{1,\dots,d\}$ in the sine block, the matrix entries are given by
\begin{align*}
    \mathcal{G}^{cc}_{pq}(f) &= \frac{C_{|p-q|} + C_{p+q}}{2}, \\
    \mathcal{G}^{ss}_{pq}(f) &= \frac{C_{|p-q|} - C_{p+q}}{2}, \\
    \mathcal{G}^{cs}_{pq}(f) &= \frac{S_{p+q} + \operatorname{sgn}(q-p)S_{|q-p|}}{2}, \\
    \mathcal{G}^{sc}(f) &= \mathcal{G}^{cs}(f)^\mathsf{T}.
\end{align*}
Since $S_0=S_0^{Y}=0$, the $q=0$ sine equation can be omitted entirely.

\subsection{Complex Toeplitz formulation}

The real-basis formulation above is adequate for low values of $d$, but it fails to exploit the symmetry of the system. We therefore change basis to the complex exponential representation of the trigonometric polynomial,
\begin{equation}
    \phi_h(t)=e^{2\pi i h f t}, \qquad h \in \{-d,\dots,d\}.
\end{equation}
Here we use the conjugated harmonic sums, which can be evaluated via NuFFT-1,
\begin{equation}
\begin{aligned}
    E_k(f) &= C_k-iS_k
    = \sum_{m=1}^{M} w_m e^{-2\pi i k f t_m}, \\
    E_k^{Y}(f) &= C_k^{Y}-iS_k^{Y}
    = \sum_{m=1}^{M} w_m y_m e^{-2\pi i k f t_m},
\end{aligned}
\end{equation}
with $E_{-k}=\overline{E_k}$ and $E_{-k}^{Y}=\overline{E_k^{Y}}$. If $\boldsymbol{x}(f)=(x_{-d},\dots,x_d)^\mathsf{T}$ denotes the coefficient vector in the complex basis, the same least-squares problem becomes
\begin{align}
    \mathcal{T}(f)\boldsymbol{x}(f)&=\boldsymbol{E}^{Y}(f), \\
    \mathcal{T}_{pq}(f)&=E_{p-q}(f), \\
    \left(\boldsymbol{E}^{Y}(f)\right)_p&=E_p^{Y}(f),
\end{align}
for $p,q \in \{-d,\dots,d\}$. In this representation, the $\mathcal{T}(f)$ matrix exhibits Hermitian Toeplitz symmetry while remaining mathematically equivalent to the real formulation.

\subsection{Levinson recursion and conditioning}
\label{sec:levinson}

Several Toeplitz-specific solvers could be applied to this system. While so-called superfast Toeplitz solvers can reach complexities as low as $\mathcal{O}(d\log^2 d)$ per system, they only become competitive at $d\gtrsim 100$ \citep{ammar1988}. Since most periodogram applications are limited to much lower degrees, we instead focus on the $\mathcal{O}(d^2)$ Levinson recursion \citep{levinson1946}, which is typically faster for the relevant problem sizes.

After shifting the harmonic indices by $d$, the Toeplitz system has size $2d+1$, with shifted right-hand side $\boldsymbol{b}(f) = (b_0,\dots,b_{2d})^\mathsf{T}$ given by $b_j \equiv E_{j-d}^{Y}(f)$ for $j=0,\dots,2d$. Let $\boldsymbol{a}^{(j)}=(a_0^{(j)},\dots,a_j^{(j)})^\mathsf{T}$ denote the coefficient vector of the predictor polynomial of order $j$, with $a_0^{(j)}=1$, and let $\varepsilon_j$ be the corresponding prediction error, initialized to $\varepsilon_0=E_0=C_0$. The Levinson recursion may then be written as
\begin{align}
    L_j &= \sum_{\ell=0}^{j-1} E_{j-\ell} a_{\ell}^{(j-1)}, \\
    K_j &= -\frac{L_j}{\varepsilon_{j-1}}, \\
    \varepsilon_j &= \varepsilon_{j-1}(1-|K_j|^2),
\end{align}
for $j=1,\dots,2d$. The quantities $K_j$ are the reflection coefficients, also known as partial correlation coefficients \citep{durbin1960}. Given these values, the predictor polynomial can be updated via
\begin{align}
    a_j^{(j)} &= K_j, \\
    a_{\ell}^{(j)} &= a_{\ell}^{(j-1)} + K_j \overline{a_{j-\ell}^{(j-1)}}, \qquad \ell=1,\dots,j-1.
\end{align}
The solution then follows from
\begin{align}
    \label{eq:levinson_residual}
    m_j &= b_j - \sum_{\ell=0}^{j-1} E_{j-\ell}x_{\ell}^{(j-1)}, &
    q_j &= \frac{m_j}{\varepsilon_j},
\end{align}
followed by
\begin{align}
    \label{eq:levinson_x_diag}
    x_j^{(j)} &= q_j, \\
    \label{eq:levinson_x_update}
    x_{\ell}^{(j)} &= x_{\ell}^{(j-1)} + q_j \overline{a_{j-\ell}^{(j)}}, \qquad \ell=0,\dots,j-1.
\end{align}

As derived by \citet{cybenko1980}, the condition number of the solved system does not exceed
\begin{equation}
    \label{conditionality}
    \kappa(\mathcal{T}) \leq \bar{\kappa}(\mathcal{T}) \equiv (2d+1) \cdot \prod_{j=1}^{2d} \frac{1+|K_j|}{1-|K_j|}.
\end{equation}
A candidate frequency is rejected as potentially ill-conditioned if $\bar{\kappa}(\mathcal{T}) > \kappa_{\mathrm{max}}$, where $\kappa_{\mathrm{max}}$ is a configurable threshold.

\subsection{Coefficient of determination}

Since the real and complex exponential formulations of the fast $\chi^2$ periodogram are equivalent, the coefficient of determination (also known as the chi-squared ratio) takes an analogous form. In both cases, the reference $\chi_0^2$ takes the value
\begin{equation}
    \chi_0^2 = YY - \frac{Y^2}{C_0}.
\end{equation}
If $\boldsymbol{x}(f)$ solves $\mathcal{T}(f)\boldsymbol{x}(f)=\boldsymbol{E}^{Y}(f)$, the minimized chi-squared at frequency $f$ is equal to
\begin{equation}
    \label{eq:chi2_min}
    \chi^2(f) = YY - \operatorname{Re}\left[\boldsymbol{E}^{Y}(f)^\ast\boldsymbol{x}(f)\right].
\end{equation}
Consequently, the coefficient of determination at a given trial frequency $f$ can be expressed as
\begin{equation}
    \label{eq:r2}
    R^2(f)=1-\frac{\chi^2(f)}{\chi_0^2}
    = \frac{\operatorname{Re}\left[\boldsymbol{E}^{Y}(f)^\ast\boldsymbol{x}(f)\right]-Y^2/C_0}{YY-Y^2/C_0},
\end{equation}
which is also known as the ``standard'' normalization of the periodogram.

Moreover, evaluating the coefficients of determination across the frequency search grid requires only the scalar quadratic form $\operatorname{Re}\left[\boldsymbol{E}^{Y}(f)^\ast\boldsymbol{x}(f)\right] = \boldsymbol{b}(f)^\ast \mathcal{T}(f)^{-1}\boldsymbol{b}(f)$, rather than the explicit coefficient vector $\boldsymbol{x}(f)$. In the Levinson recursion, this quantity can be accumulated directly during the forward pass as
\begin{equation}
    \label{eq:r2_forward}
    \operatorname{Re}\left[\boldsymbol{E}^{Y}(f)^\ast\boldsymbol{x}(f)\right] = \sum_{j=0}^{2d} \frac{|\xi_j|^2}{\varepsilon_j},
\end{equation}
where $\xi_0 = b_0$ and
\begin{equation}
    \xi_j = \sum_{\ell=0}^{j} a_{\ell}^{(j)} b_{j-\ell}, \qquad j=1,\dots,2d.
\end{equation}
Because the predictor coefficients $a_{\ell}^{(j)}$ and prediction errors $\varepsilon_j$ are already evaluated during the forward pass of the Levinson recursion, Eq.~\eqref{eq:r2_forward} determines $R^2(f)$ without computing the coefficient vector $\boldsymbol{x}(f)$. Consequently, if the exact values of the fitted harmonic coefficients are not required across the search grid, the solution recurrence given by Eqs.~\eqref{eq:levinson_residual}--\eqref{eq:levinson_x_update} can be omitted entirely, avoiding the residual evaluations and solution updates at each trial frequency.

\section{Comparison with existing implementations}
\label{sec:comparison}

To directly compare our modified algorithm with currently available open-source software, we implemented the algorithms described in Sects.~\ref{sec:emp_complexity} and~\ref{sec:linear_systems}. The implementation relies on the decimation-in-frequency fast Fourier transform \citep{dif_fft} and the COBRAVO bit-reversal permutation algorithm \citep{COBRAVO}, using single-, double-, and double-double-precision arithmetic \citep{Dekker_Double_Double}. For the spreading stage, in which the samples are projected onto a uniform grid by convolution with a finite-width kernel, we rely on the zeroth-order prolate spheroidal wave function \citep{Osipov2013ProlateSW}, approximated by a modified Remez algorithm \citep{minimax_approx}, and Gauss--Legendre quadrature for deconvolution \citep{finufft}. We modified our implementations to ignore frequencies at which the upper bound on the condition number, estimated according to Eq.~\eqref{conditionality}, exceeds $5\times 10^{5}$ for the single-precision implementation and $10^{13}$ for the double-precision implementation, to ensure numerical stability.

Due to the complexity of the described algorithm, we limit ourselves to a high-level description of the implementation. Readers interested in further details can find the code used to obtain the results presented below in the publicly available Toeplitz-LS repository.

\subsection{Tests on synthetic white noise}
\begin{figure}
  \resizebox{\hsize}{!}{\includegraphics{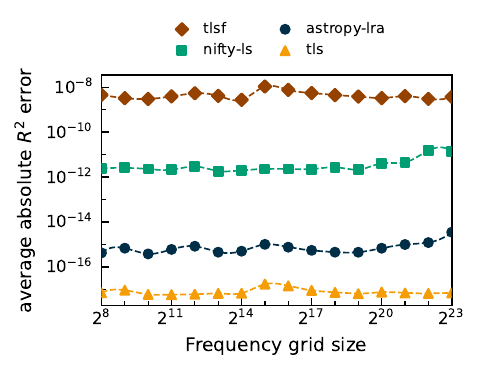}}
  \caption{Average absolute $R^2$ error as a function of frequency grid size for degree-one periodograms, as measured for a white noise synthetic light curve containing $M=1000$ measurements.}
  \label{fig:synthetic_error}
\end{figure}

To validate the reference implementation, we generated artificial light curves filled with white noise, each containing $M=1000$ measurements. We computed $R^2$ values for grid sizes between $N=2^8$ and $N=2^{23}$ target frequencies in the $d=1$ case, which is mathematically equivalent to the generalized Lomb--Scargle periodogram \citep{1981AJ.....86..619F, GLS}. We used our double-double implementation as the reference and computed average differences over ten runs for our single-precision (\textsc{tlsf}) and double-precision (\textsc{tls}) implementations. For comparison, we included \textsc{nifty-ls} \citep{Garrison_2024}, using version 1.1.0\footnote{\url{https://github.com/flatironinstitute/nifty-ls}} throughout all benchmarks, and \textsc{astropy-lra} \citep{astropy}, available in Astropy from version 8.0\footnote{\url{https://docs.astropy.org/en/stable/changelog.html\#version-8-0-0-2026-06-16}}. The full results are shown in Fig.~\ref{fig:synthetic_error}.

Our accuracy estimates for \textsc{nifty-ls} match those reported by \citet{Garrison_2024} within a single order of magnitude. We attribute the 12-order-of-magnitude increase in the accuracy of the Astropy package, relative to the values reported for the older \textsc{astropy-fasper} implementation, to the introduction of the new \textsc{astropy-lra} implementation based on the low-rank approximation \citep{LRA_NFFT}. In contrast, \textsc{astropy-fasper} relies on time-domain Lagrange polynomial interpolation following \citet{Press1989}, leading to a significant divergence from the numerically accurate values. Our double-precision implementation exceeds the accuracy of \textsc{nifty-ls} and \textsc{astropy-lra} across the entire frequency grid size range. Additionally, the single-precision implementation retains an average error below $10^{-8}$ for all of the tested grid sizes, making it more than adequate for most practical applications.

\begin{figure}
  \resizebox{\hsize}{!}{\includegraphics{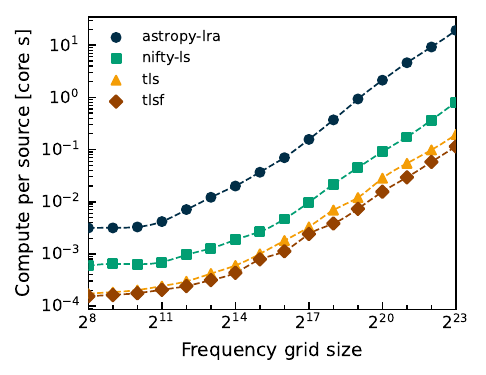}}
  \caption{Computation time as a function of frequency grid size for degree-one synthetic benchmarks.}
  \label{fig:synthetic_runtime}
\end{figure}

In addition to confirming the accuracy of our reference implementation, we measured the performance of all tested methods as a function of the grid length. All measurements were performed on a 16-core AMD Ryzen~9 7945HX CPU; the C code was compiled with version 14.2 of the GNU Compiler Collection on a Linux-based operating system with a Python 3.13 runtime. In the synthetic tests, the execution times for both of our implementations scale roughly proportionally to those of \textsc{nifty-ls}, with the double-precision implementation at least twice as fast as \textsc{nifty-ls}. The single-precision implementation consistently produces shorter runtimes than the double-precision implementation, with a maximum speed-up factor of $4.15\times$ at $N=2^{23}$. The \textsc{nifty-ls} package outperforms \textsc{astropy-lra} at all tested grid sizes, with a gap exceeding that between our implementations and \textsc{nifty-ls}. The full results are shown in Fig.~\ref{fig:synthetic_runtime}.

\begin{figure}
  \resizebox{\hsize}{!}{\includegraphics{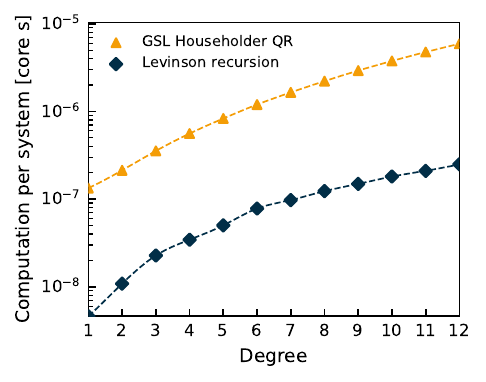}}
  \caption{Computation time per system as a function of the degree of the trigonometric polynomial for the Levinson recursion and the GNU Scientific Library Householder QR solver.}
  \label{fig:levinson_gsl}
\end{figure}

In addition to the $d=1$ runs, we measured the average time needed to solve the system of linear equations used by the periodogram at each frequency, for $1 \leq d \leq 12$. We compared the Levinson recursion against the Householder QR decomposition provided by the GNU Scientific Library \citep{gsl} and used in the reference implementation of \citet{palmer2009}. The Levinson recursion achieves a minimum speed-up factor of $15.35\times$ at $d=6$, with the largest speed-up factors at $d=1$ ($29.0\times$) and $d=12$ ($23.7\times$). The full results are shown in Fig.~\ref{fig:levinson_gsl}.

\subsection{Tests on OGLE-IV variable-star photometry}

Synthetic data benchmarks provide a useful baseline, but they are not representative of large-scale time-domain sky surveys. There are two key differences that our synthetic tests did not cover. First, observational gaps and visibility windows introduce a risk of ill-conditioning in the normal equations \citep{garcia2017}. Second, multi-core CPU architectures exhibit parallel scaling constraints; total CPU time (measured as cumulative core-hours across all active cores) often increases significantly with the number of worker threads due to memory-bandwidth saturation, as noted by \citet{Garrison_2024}.

To evaluate the performance of our algorithm under realistic observational conditions, we benchmarked our \textsc{tlsf} and \textsc{tls} implementations against five existing implementations: \textsc{astropy-lra} and \textsc{astropy-fasper}, provided as core components of the Astropy library \citep{astropy}; \textsc{aovdist}\footnote{\label{fn:aovdist}\url{http://users.camk.edu.pl/alex/soft/aovdist.tgz}}, a reference implementation of the Szeg\H{o} recursion-based algorithm \citep{schwartzenberg-czerny1996}; \textsc{fastchi}\scalebox{0.75}{2}~\textsc{v}\scalebox{0.75}{1.03}\footnote{\url{https://web.archive.org/web/20250525051459/http://public.lanl.gov/palmer/fastchi/FastChi2-1.03.tar.gz}}, a reference implementation of the algorithm published by \citet{palmer2009}; and \textsc{nifty-ls}. All observational light curves were obtained by the Optical Gravitational Lensing Experiment (OGLE-IV; \citealt{OGLE-IV}), specifically focusing on two distinct variable-star sub-catalogs from the Large Magellanic Cloud (LMC): pulsating RR~Lyrae stars \citep{LMC-RRLyr} and eclipsing binary systems \citep{LMC-ECL}.

All benchmarks were evaluated on identical uniform frequency grids for each light curve, starting at $f_{\mathrm{min}} = 0$ up to a dataset-specific maximum frequency $f_{\mathrm{max}}$, with uniform frequency spacing $\Delta_f = 1/(5 d \Delta_t)$, where $\Delta_t = \max(t_m) - \min(t_m)$ is the observational time baseline of the light curve and the factor of $5d$ provides five samples per highest-harmonic resolution element $1/(d\Delta_t)$. A best-fit frequency was classified as correct if its absolute difference from the catalog frequency did not exceed $1 / \Delta_t$. All computations were performed on the 16-core AMD Ryzen~9 7945HX CPU described above, using 15 parallel threads for all tested implementations, with each thread independently processing one lightcurve at a time.

\subsubsection{OGLE-IV LMC RR~Lyrae}
As the first reference dataset, we analyzed 39\,931 light curves of OGLE-IV LMC RR~Lyrae stars containing at least 50 measurements each (with an average of 492 measurements per light curve). As an external reference, we benchmarked \textsc{fnpeaks}\footnote{\url{http://helas.astro.uni.wroc.pl/deliverables.php?lang=en&active=fnpeaks}}, which estimates periods using a weighted Rayleigh $Z$ test \citep[Sect.~6.2.1b]{Mardia1972}, using a fixed frequency step of $\Delta_f = 5\times 10^{-5}\;\mathrm{d}^{-1}$ (cycles per day) up to $f_{\mathrm{max}}=5\;\mathrm{d}^{-1}$ and obtaining a recovery rate of 98.20\% in 0.098~core-hours\footnote{\label{fn:fnpeaks_flags} Compiled with the default \texttt{-msse} flag replaced by \texttt{-march=native}; with the default flag, execution required 0.17~core-hours on the RR~Lyrae dataset and 0.13~core-hours on the eclipsing binary dataset at the same recovery rates.}.

For the least-squares implementations, we generated power spectra for each target for both the single-harmonic case ($d=1$, corresponding to the generalized Lomb--Scargle periodogram) and $d=3$, representative of a typical degree used for period determination of pulsating variables, evaluated up to $f_{\mathrm{max}}=5\;\mathrm{d}^{-1}$. The \textsc{astropy-lra}, \textsc{aovdist}, \textsc{nifty-ls}, and \textsc{tls} implementations (along with \textsc{tlsf} at $d=1$) all returned identical best-fit frequencies across all light curves tested, achieving period recovery rates of 98.91\% at $d=1$ and 96.97\% at $d=3$. In contrast, \textsc{astropy-fasper} and \textsc{fastchi}\scalebox{0.75}{2}~\textsc{v}\scalebox{0.75}{1.03} returned different periods for a subset of the tested stars in both benchmarks, most likely as a result of their less accurate underlying NuFFT-1 approximations.

Perhaps counterintuitively, at $d=3$, \textsc{tlsf} returned slightly more best-fit frequencies matching catalog values than the more numerically accurate implementations. In those light curves, the peak frequencies identified by the more precise methods lay in regions where the estimated condition number $\bar{\kappa}$ exceeded the $\kappa_{\mathrm{max}} = 5\times 10^5$ limit, which in the case of \textsc{tlsf} likely reduced its susceptibility to aliasing. We discuss these findings in more detail in Sect.~\ref{sec:discussion}.

In addition to matching or exceeding the recovery rates of the other tested implementations, \textsc{tlsf} achieved the lowest computational cost in both benchmarks. At $d=1$, \textsc{tlsf} processed the entire catalog at a total computational cost of $8$ core-seconds, corresponding to a throughput of over $75\,000$ light curves per second and a speed-up of nearly $30\times$ over state-of-the-art \textsc{nifty-ls} and over $3500\times$ over the multithreaded \textsc{astropy-lra} implementation. At $d=3$, \textsc{tlsf} executed the benchmark at a total computational cost of $0.054$ core-hours, representing a speed-up of over $2600\times$ over \textsc{astropy-lra} and more than $26\times$ over the \textsc{nifty-ls} library. The full benchmark results are summarized in Tables~\ref{tab:rrlyr_d1} and~\ref{tab:rrlyr_d3}.

\begin{table}
\caption{Period recovery rates and computational costs for the OGLE-IV LMC RR~Lyrae dataset at $d=1$.}
\label{tab:rrlyr_d1}
\centering
\small
\begin{tabular}{l l l}
\hline\hline
Implementation & Recovery rate & Computation time \\
               & (\%)          & (core-h) \\
\hline
\textsc{astropy-lra}    & 98.91 & 7.788 \\
\textsc{astropy-fasper} & 98.86 & 3.437 \\
\textsc{aovdist}        & 98.91 & 1.051 \\
\textsc{fastchi}\scalebox{0.75}{2}~\textsc{v}\scalebox{0.75}{1.03} & 98.72 & 0.310 \\
\textsc{nifty-ls}       & 98.91 & 0.0650 \\
\hline
\textsc{tls}            & 98.91 & 0.0062 \\
\textsc{tlsf}           & 98.91 & 0.0022 \\
\hline
\end{tabular}
\tablefoot{Light curves with fewer than 50 measurements were excluded, yielding 39\,931 light curves (averaging 492 measurements per curve) evaluated up to $f_{\mathrm{max}} = 5\;\mathrm{d}^{-1}$.}
\end{table}

\begin{table}
\caption{Period recovery rates and computational costs for the OGLE-IV LMC RR~Lyrae dataset at $d=3$.}
\label{tab:rrlyr_d3}
\centering
\small
\begin{tabular}{l l l}
\hline\hline
Implementation & Recovery rate & Computation time \\
               & (\%)          & (core-h) \\
\hline
\textsc{astropy-lra}    & 96.97 & 144.492 \\
\textsc{astropy-fasper} & 95.22 & 138.588 \\
\textsc{aovdist}        & 96.97 & 5.306 \\
\textsc{fastchi}\scalebox{0.75}{2}~\textsc{v}\scalebox{0.75}{1.03} & 94.80 & 5.178 \\
\textsc{nifty-ls}       & 96.97 & 1.413 \\
\hline
\textsc{tls}            & 96.97 & 0.165 \\
\textsc{tlsf}           & 96.98 & 0.054 \\
\hline
\end{tabular}
\tablefoot{Light curves with fewer than 50 measurements were excluded, yielding 39\,931 light curves (averaging 492 measurements per curve) evaluated up to $f_{\mathrm{max}} = 5\;\mathrm{d}^{-1}$.}
\end{table}

\subsubsection{OGLE-IV LMC eclipsing binaries}

In trigonometric least-squares periodograms, the sharper and more complex variability profiles of eclipsing binaries benefit from a larger number of harmonics to a significantly greater extent than those of pulsating variables. We therefore benchmarked the algorithms across 38\,002 OGLE-IV LMC eclipsing binary light curves \citep{LMC-ECL} containing at least 50 measurements (with an average of 509 measurements per light curve), evaluating frequency grids up to $f_{\mathrm{max}}=4\;\mathrm{d}^{-1}$. Because the optimal degree of the fitted trigonometric polynomial varies substantially across different datasets, we initialized the search at $d=1$ and progressively increased $d$ until an execution budget of 30~core-hours per algorithm was exhausted or the algorithm completed the pass at $d=12$. Furthermore, due to the uncertainty in detecting secondary eclipses in noisy light curves, we classified a period as successfully recovered if the identified period matched the catalog period, half of it ($0.5\times$), or twice it ($2.0\times$), where the match was assessed in frequency space with the same $1 / \Delta_t$ tolerance. Using the same fixed frequency step of $\Delta_f = 5\times 10^{-5}\;\mathrm{d}^{-1}$, \textsc{fnpeaks} achieved a recovery rate of 72.94\% in 0.079~core-hours\textsuperscript{\ref{fn:fnpeaks_flags}}.

\begin{figure}
  \resizebox{\hsize}{!}{\includegraphics{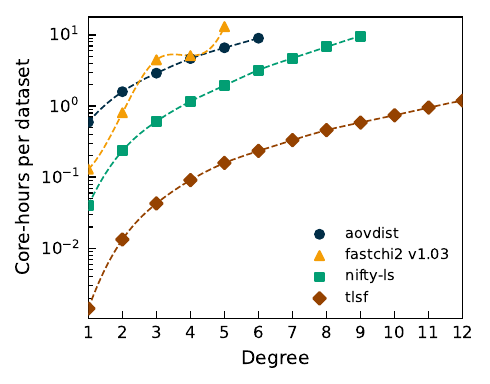}}
  \caption{Computation time as a function of trigonometric-polynomial degree for the OGLE-IV LMC eclipsing-binary benchmark.}
  \label{fig:ecl_sweep_time}
\end{figure}

\begin{figure}
  \resizebox{\hsize}{!}{\includegraphics{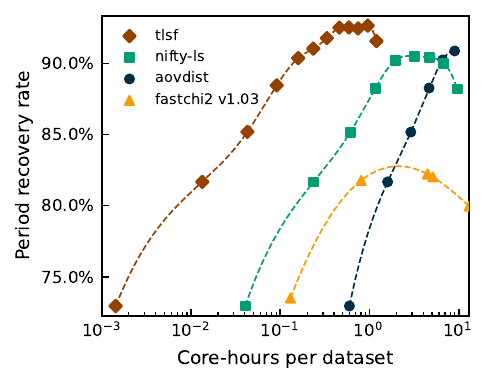}}
  \caption{Period recovery rate as a function of computation time for the OGLE-IV LMC eclipsing-binary benchmark.}
  \label{fig:ecl_sweep_accuracy}
\end{figure}

Figure~\ref{fig:ecl_sweep_time} shows the measured computation time as a function of trigonometric-polynomial degree, and Fig.~\ref{fig:ecl_sweep_accuracy} illustrates the period recovery rate as a function of computation time (with \textsc{tls} omitted from both figures for visual clarity). The \textsc{tls} and \textsc{tlsf} solvers were the only implementations to reach the degree cap of $d=12$ within the specified 30~core-hour cumulative computational budget. Furthermore, \textsc{tlsf} was both the least computationally expensive and the most accurate method, achieving the highest period recovery rate of 92.66\% at a computational cost of 0.93~core-hours.

In addition to the multi-degree sweep, we performed a dedicated benchmark at $d=11$ to compare the tested implementations under an equivalent problem formulation. We excluded the Astropy implementations from this comparison due to their prohibitively high computational cost, exceeding $10^4$~core-hours under their expected $\mathcal{O}(d^4)$ computational complexity within the specified problem formulation. The benchmark results are summarized in Table~\ref{tab:ecl_d11}. Unlike the pulsating variable test, the recovery rates for eclipsing binaries diverged widely among the implementations as a result of design matrix ill-conditioning for high-degree trigonometric polynomials. In particular, \textsc{fastchi}\scalebox{0.75}{2}~\textsc{v}\scalebox{0.75}{1.03} lacks any condition-number check, accepting unconstrained solutions to ill-conditioned systems and suffering severe numerical degradation, especially when combined with a low-accuracy NuFFT-1 implementation (recovering only 33.86\% of periods). Under these conditions, \textsc{tlsf} remained both the most computationally efficient and most accurate of the tested implementations, retaining a speed-up of more than $16\times$ compared to \textsc{nifty-ls}.

We attribute the instability in period determination most likely to the markedly different frequency-rejection strategies of the implementations. Specifically, \textsc{nifty-ls} aborted the period search whenever at least one trial frequency was detected as potentially unstable, which occurred for over 10\% of the dataset, whereas \textsc{aovdist} rejected only the harmonics determined to be potentially unstable, leaving occasional detections of false periods caused by numerical error propagation.

\begin{table}
\caption{Period recovery rates and computational costs for the OGLE-IV LMC eclipsing binary dataset at $d=11$.}
\label{tab:ecl_d11}
\centering
\small
\begin{tabular}{l l l}
\hline\hline
Implementation & Recovery rate & Computation time \\
               & (\%)          & (core-h) \\
\hline
\textsc{fastchi}\scalebox{0.75}{2}~\textsc{v}\scalebox{0.75}{1.03} & 33.86 & 142.593 \\
\textsc{aovdist}        & 90.70 & 26.547 \\
\textsc{nifty-ls}       & 84.12 & 16.071 \\
\hline
\textsc{tls}            & 92.56 & 1.998 \\
\textsc{tlsf}           & 92.66 & 0.928 \\
\hline
\end{tabular}
\tablefoot{Light curves with fewer than 50 measurements were excluded, yielding 38\,002 light curves (averaging 509 measurements per curve) evaluated up to $f_{\mathrm{max}} = 4\;\mathrm{d}^{-1}$.}
\end{table}

\section{Discussion}
\label{sec:discussion}

Multiharmonic least-squares periodograms provide an effective method for detecting and characterizing non-sinusoidal periodic variability in unevenly sampled astronomical time series. In large-scale sky surveys, however, the computational cost of evaluating trigonometric polynomials of degree $d > 1$ often leads pipelines to adopt the single-harmonic case, also known as the generalized Lomb--Scargle periodogram \citep{1981AJ.....86..619F, GLS}. Classical methods either scale unfavorably with $M$ (such as the $\mathcal{O}(dNM)$ Szeg\H{o} recursion) or rely on general-purpose linear solvers that incur significant computational overhead at higher degrees as a result of their $\mathcal{O}(d^3)$ scaling.

In this work, we explored whether a multiharmonic trigonometric periodogram can achieve computational throughput and cache efficiency comparable to single-harmonic algorithms without sacrificing period recovery accuracy. By splitting the large-grid NuFFT-1 into multiple cache-resident blocks and solving the resulting complex Hermitian Toeplitz systems via the $\mathcal{O}(d^2)$ Levinson recursion, both the Fourier evaluation and linear solve stages become substantially more efficient. The overall computational complexity scales as $\mathcal{O}(dN(\log M + d))$, bringing high-degree evaluations within practical reach.

Our benchmark results confirm the practical advantages of this formulation across synthetic datasets and large observational catalogs. On synthetic white noise (Figs.~\ref{fig:synthetic_error}--\ref{fig:levinson_gsl}), \textsc{tls} delivers higher accuracy than existing tools while running more than twice as fast as \textsc{nifty-ls} for $d=1$. The \textsc{tlsf} solver provides an additional speed-up, maintaining average $R^2$ differences below $10^{-8}$ in our benchmarks with $M=1000$ measurements (noting that this error depends on $M$). Across tens of thousands of OGLE-IV light curves of pulsating RR~Lyrae stars ($d=3$, Table~\ref{tab:rrlyr_d3}) and eclipsing binary systems ($d=11$, Table~\ref{tab:ecl_d11}, Figs.~\ref{fig:ecl_sweep_time} and~\ref{fig:ecl_sweep_accuracy}), our implementations achieve the highest period recovery rates while reducing execution times by over an order of magnitude compared to previously available implementations.

\subsection{Effects of matrix conditioning on recovery rates}
\label{sec:conditionality}

\begin{figure}
  \resizebox{\hsize}{!}{\includegraphics{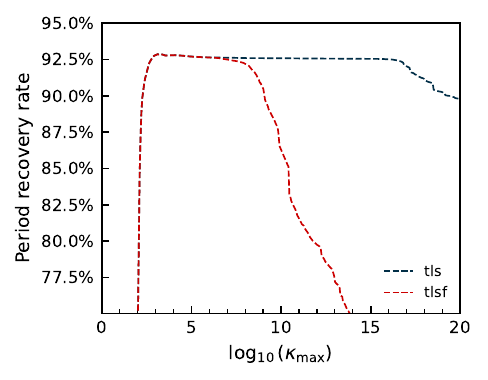}}
  \caption{Period recovery rate of single-precision (\textsc{tlsf}) and double-precision (\textsc{tls}) implementations as a function of the conditioning threshold $\kappa_{\mathrm{max}}$ for the OGLE-IV LMC eclipsing binary dataset at $d=11$.}
  \label{fig:condition}
\end{figure}

\begin{figure}
  \resizebox{\hsize}{!}{\includegraphics{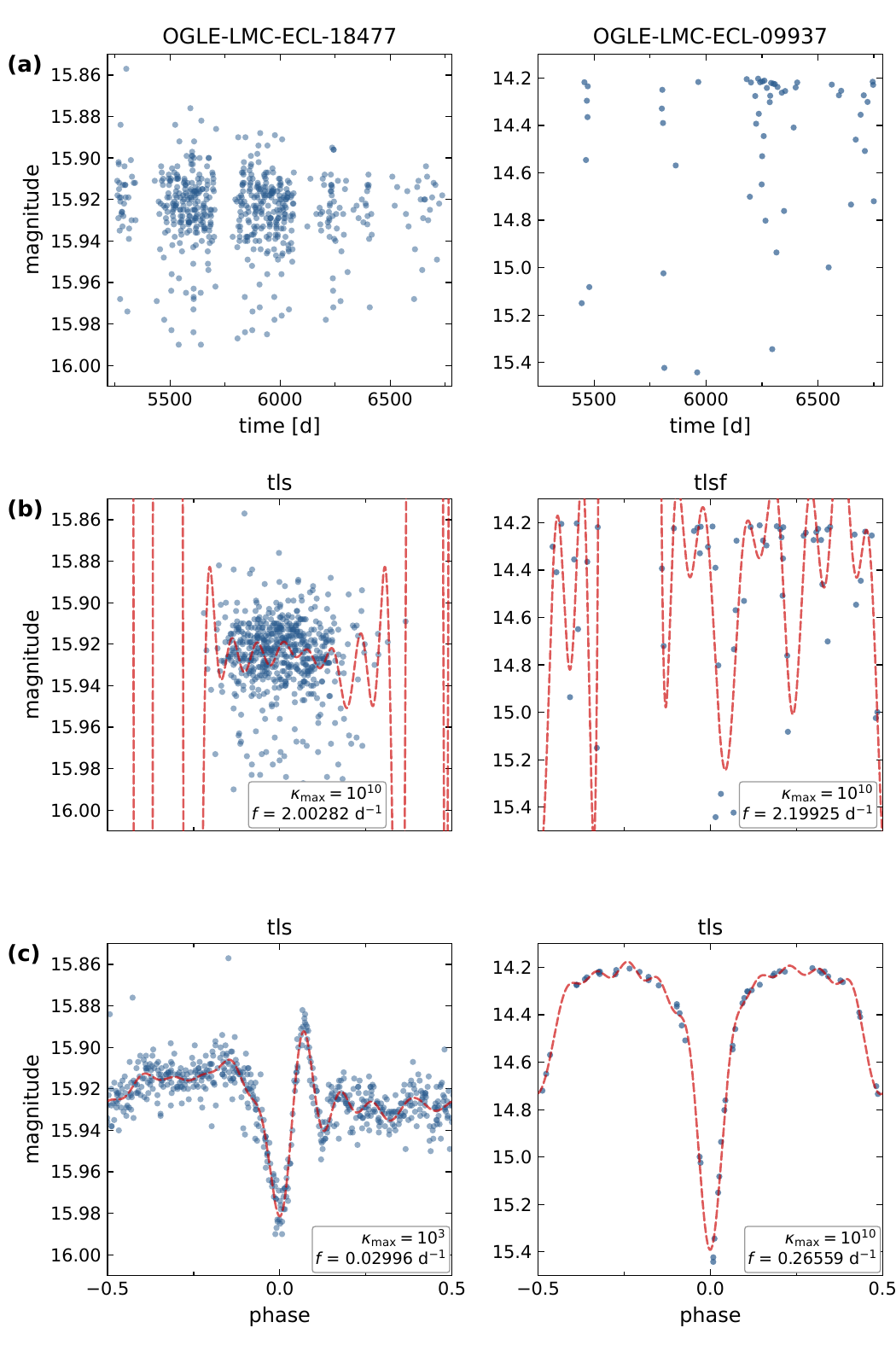}}
  \caption{Light curves of OGLE-LMC-ECL-18477 (left) and OGLE-LMC-ECL-09937 (right) with fitted $d=11$ models (dashed curves). The rows show (a) the raw light curves with observation times in $\mathrm{HJD} - 2\,450\,000$, measured in days ($\mathrm{d}$), (b) misphased fits, and (c) fits recovering the catalog periods, with zero phase selected at the moment of minimum brightness.}
  \label{fig:instability_alias}
\end{figure}

In multiharmonic least-squares fitting, observational cadence gaps and annual visibility windows frequently degrade the conditioning of the normal equations. For $d=11$ on the OGLE-IV LMC eclipsing binary dataset, the \textsc{tlsf} and \textsc{tls} implementations achieve nearly identical period recovery rates for conditioning thresholds $\kappa_{\mathrm{max}} \lesssim 10^{6}$ (Fig.~\ref{fig:condition}). Above that threshold, the recovery rate of \textsc{tlsf} deteriorates as $\kappa_{\mathrm{max}} \cdot \epsilon$ approaches unity (where $\epsilon$ denotes the relative numerical accuracy of the underlying NuFFT-1 approximation), whereas \textsc{tls} remains reliable up to $\kappa_{\mathrm{max}} \sim 10^{16}$. This precision-dependent instability is illustrated in Fig.~\ref{fig:instability_alias} (right column) for OGLE-LMC-ECL-09937 at $\kappa_{\mathrm{max}} = 10^{10}$, where \textsc{tls} recovers the true period ($f = 0.26559\;\mathrm{d}^{-1}$) while \textsc{tlsf} fails on a spurious peak ($f = 2.19925\;\mathrm{d}^{-1}$) caused by numerical round-off amplification.

Because both implementations reach their optimal period recovery rate at $\kappa_{\mathrm{max}} \lesssim 10^{3}$, single precision is expected to be sufficient for most survey pipelines when coupled with a suitable condition-number cutoff. Furthermore, this conditioning behavior is the likely cause of \textsc{tlsf} consistently exceeding the recovery rate of \textsc{tls} across both the RR~Lyrae and eclipsing binary benchmarks (Tables~\ref{tab:rrlyr_d3} and~\ref{tab:ecl_d11}) by a small margin. The lower rejection threshold in \textsc{tlsf} acts as an effective regularizer, rejecting peaks with weak phase coverage that otherwise mislead unconstrained solvers. Figure~\ref{fig:instability_alias} (left column) illustrates this regime for OGLE-LMC-ECL-18477: at $\kappa_{\mathrm{max}} = 10^{10}$, well within the numerically stable range of double precision, \textsc{tls} locks onto a strong nightly alias ($f = 2.00282\;\mathrm{d}^{-1}$), whereas a stricter cutoff of $\kappa_{\mathrm{max}} = 10^{3}$ rejects the ill-conditioned alias and recovers the true period ($f = 0.02996\;\mathrm{d}^{-1}$). We also note that exploring the feasibility of soft regularization of the periodogram test statistic via $\bar{\kappa}$ could potentially reduce the susceptibility of trigonometric periodograms to aliasing.

It should also be emphasized that the reported period recovery rates are not directly representative of periodogram accuracy in survey pipelines, where more than a single peak is usually evaluated and most instability-induced false peaks can be automatically discarded as they often yield nonphysical $R^2 > 1$ values. Even so, these numbers offer valuable insight into the numerical stability of different implementations under typical survey conditions.

\subsection{Numerical error propagation on equispaced frequency grids}
\label{sec:error}

In our synthetic benchmarks (Fig.~\ref{fig:synthetic_error}), the measured average $R^2$ error remains approximately constant across grid sizes up to $N = 2^{23}$, well below the $\mathcal{O}(N \epsilon_{\mathrm{mach}})$ lower bound on the achievable error proposed by \citet{Garrison_2024}, where $\epsilon_{\mathrm{mach}}$ is the upper bound on the relative approximation error due to rounding in the floating-point format used. As shown in Fig.~\ref{fig:synthetic_error}, neither our implementations nor \textsc{nifty-ls} exhibit the linear $\mathcal{O}(N \epsilon_{\mathrm{mach}})$ error accumulation that \citet{Garrison_2024} suggested would render single-precision periodograms unusable at large grid sizes. Following the numerical stability analysis of \citet{cybenko1980}, we bound the worst-case error accumulation of the Levinson recursion applied to Hermitian Toeplitz systems by $\mathcal{O}(d^2 \cdot \epsilon \cdot \kappa_{\mathrm{max}} \cdot \log N)$.

Additionally, this upper bound can be lowered to $\mathcal{O}(\kappa_{2} \cdot \epsilon \cdot \log N)$, with an average accumulated error of $\mathcal{O}(\kappa_{2} \cdot \epsilon \cdot \sqrt{\log N})$, if the Levinson recursion is replaced by a more numerically stable linear solver such as the singular value decomposition \citep[Chaps.~20 and 21]{Higham2002}, where $\kappa_2 \le \bar{\kappa} \le \kappa_{\mathrm{max}}$. Moreover, our empirical tests show that single precision preserves high recovery rates even at degrees as high as $d=11$ (Fig.~\ref{fig:condition}). Consequently, half-precision implementations may be viable for initial candidate period screening. While our theoretical results show that error accumulation scales favorably as $\mathcal{O}(\sqrt{\log N})$ on equispaced grids, determining rigorous lower bounds on both the worst-case error and the average error growth remains an open problem.

\subsection{Multithreading and scalability}
\label{sec:memory_scaling}

Our benchmarks also reveal a pronounced difference in how multithreading affects the performance of different implementations. In our tests, \textsc{fastchi}\scalebox{0.75}{2}~\textsc{v}\scalebox{0.75}{1.03} and \textsc{nifty-ls} exhibited strongly sublinear speed-ups with thread count, consistent with the memory-bandwidth bottleneck identified by \citet{Garrison_2024}. In contrast, our implementations and \textsc{aovdist} achieved near-linear scaling up to the 16 physical cores available on our test platform.

\begin{figure}
  \resizebox{\hsize}{!}{\includegraphics{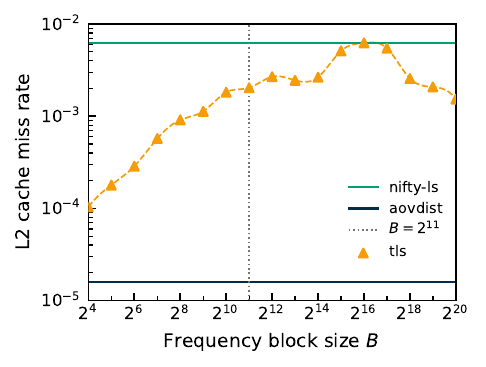}}
  \caption{L2 data-cache miss rate (defined as the total number of L2 data-cache misses divided by the total number of retired instructions) as a function of FFT block size $B=2^{k}$ for the degree-one periodogram with $M=1000$ measurements and $N=2^{20}$ frequencies. The vertical dotted line marks the native default block size $B=2^{11}$.}
  \label{fig:l2cache}
\end{figure}

Profiling the different implementations confirms that the near-linear scaling of our method stems from differences in cache locality (Fig.~\ref{fig:l2cache}). Specifically, the block-size optimization described in Sect.~\ref{sec:emp_complexity} keeps intermediate FFT results within the core-private L2 cache (yielding an L2 miss rate of $2.5\times 10^{-3}$ per instruction at the default block size $B=2^{11}$), whereas monolithic grid implementations such as \textsc{nifty-ls} exceed cache capacity ($5.9\times 10^{-3}$ misses per instruction) and incur severe DRAM memory-bandwidth contention. Furthermore, we note that, unlike the other implementations, \textsc{aovdist} recomputes the trigonometric functions at every target frequency, which keeps its working set small ($2\times 10^{-5}$ misses per instruction) at the cost of high arithmetic overhead; on equispaced frequency grids, these evaluations could be replaced by recursive complex multiplications \citep{Ponman1981, Kurtz1985}, which is likely to substantially accelerate the Szeg\H{o} recursion and potentially make it competitive for small $M$.

This cache-locality advantage of the blocked NuFFT-1 scheme is also likely the primary driver of the speed-up in our generalized Lomb--Scargle periodogram implementation (Table~\ref{tab:rrlyr_d1}), which does not rely on the Levinson recursion; this also explains why the multithreaded gain significantly exceeds the single-threaded speed-up (Fig.~\ref{fig:synthetic_runtime}). Two additional design choices further widen this margin: more extensive vectorization, and computing the $R^2(f)$ grid directly in bit-reversed order through the decimation-in-frequency FFT, which requires only a single permutation of the real-valued $R^2(f)$ array per block rather than $3d$ permutations of complex-valued FFT output grids applied by the other FFT-based implementations.

\subsection{Viability of GPU offloading}
\label{sec:cpu_gpu}

When evaluated on 15 threads of an AMD Ryzen~9 7945HX CPU, both \textsc{tls} and \textsc{tlsf} achieve higher throughput than that measured on the NVIDIA A100 GPU by \citet{Garrison_2024}. While our blocked Toeplitz algorithm is mathematically suited to GPU execution, preliminary benchmarks indicate modest speed-ups over CPU execution. Unless our algorithm is further optimized for GPU offloading, under comparable power and cost constraints, CPUs are likely to exceed the throughput measured on GPUs.

This behavior is most likely driven by the substantial dynamic memory footprint required per thread to maintain intermediate FFT results across thousands of concurrent GPU threads, which limits effective occupancy. Consequently, multi-core CPUs with large caches are likely to remain a highly practical and cost-effective hardware platform for multiharmonic period searches.

\subsection{Implications for large-scale sky surveys}
\label{sec:implications}

Our benchmarks demonstrate consistent speed-ups exceeding an order of magnitude over existing state-of-the-art tools, enabling high-degree trigonometric polynomial fitting to be deployed routinely across large-scale survey catalogs. This capability makes high-degree harmonic period searches more feasible at the scale of next-generation photometric sky surveys, such as the Vera C. Rubin Observatory Legacy Survey of Space and Time \citep[LSST;][]{LSST}. Furthermore, as a result of this improved efficiency and numerical stability, high-degree trigonometric periodograms may provide a practical alternative to specialized transit detection algorithms such as the fast-folding box least-squares algorithm \citep{fbls}.

\section{Conclusions}
\label{sec:conclusions}

We present a new implementation of the multiharmonic least-squares periodogram that is more than an order of magnitude faster than the state-of-the-art \textsc{nifty-ls} library \citep{Garrison_2024} across multithreaded benchmarks (Tables~\ref{tab:rrlyr_d1}--\ref{tab:ecl_d11}; Fig.~\ref{fig:ecl_sweep_time}). For low to moderate numbers of harmonics ($1 \le d \le 5$), our implementation matches the period recovery rates of \textsc{nifty-ls} on both RR~Lyrae (Tables~\ref{tab:rrlyr_d1} and~\ref{tab:rrlyr_d3}) and eclipsing binary light curves (Fig.~\ref{fig:ecl_sweep_accuracy}), while achieving significantly higher recovery rates at the high numbers of harmonics ($d > 5$) beneficial for sharp eclipsing binary profiles, as ill-conditioned trial frequencies become more common (Fig.~\ref{fig:ecl_sweep_accuracy}; Table~\ref{tab:ecl_d11}).

These performance improvements stem from expressing the system of normal equations as a Hermitian Toeplitz matrix, combining the AoVMH formulation of \citet{schwartzenberg-czerny1996} with the NuFFT-1 evaluation of the \textsc{fastchi}\scalebox{0.75}{2} algorithm \citep{palmer2009}. Coupled with the cache-resident blocked NuFFT-1 scheme (Sect.~\ref{sec:emp_complexity}), this representation enables $\mathcal{O}(d^2)$ model fitting via the Levinson recursion and allows the coefficient of determination $R^2(f)$ to be accumulated directly during the forward pass. Although such a system also admits superfast Toeplitz solvers of $\mathcal{O}(d\log^2 d)$ complexity, their substantial algorithmic overhead and constant factors cause them to underperform compared to the cache-friendly Levinson recursion for the harmonic degrees ($d \le 12$) typical of astronomical period searches. Furthermore, excluding individual trial frequencies at which the condition-number upper bound $\bar{\kappa}$ exceeds the cutoff $\kappa_{\mathrm{max}}$ suppresses numerical instability at high harmonic degrees, with minimal risk of rejecting true variability periods and at negligible additional cost.

Despite these encouraging results, several limitations and open questions remain. First, while the condition-number cutoff $\kappa_{\mathrm{max}}$ reduces susceptibility to aliasing, incorporating $\bar{\kappa}$ directly into the test statistic as an indicator of phase coverage may prove superior and eliminate the need for an explicit rejection threshold. Second, existing periodogram implementations based on the Szeg\H{o} recursion do not use recursive complex multiplication to avoid repeated trigonometric evaluations on equispaced frequency grids; determining the empirical crossover point in $M$ and $d$ between the Toeplitz formulation and an optimized Szeg\H{o} recursion would clarify the optimal regime for sparsely sampled light curves. Finally, while direct translation of the Levinson recursion to GPUs is currently constrained by its dynamic memory footprint, evolving accelerator architectures warrant a deeper examination of GPU and heterogeneous implementations. Ultimately, our formulation increases the practical feasibility of high-degree multiharmonic period searches on datasets from large-scale sky surveys, potentially making these searches competitive with domain-specific algorithms not included in our comparison.

\section*{Data availability}
All performance benchmarks and period recovery rates presented in this paper were derived using the Linux \texttt{x86-64} build of the \texttt{v1.0.0} release of the \texttt{Toeplitz-LS} library, which is publicly available along with the benchmark and plotting scripts at \url{https://github.com/kkrutkowski/Toeplitz-LS}; comparisons against \textsc{nifty-ls} used the \texttt{v1.1.0} source distribution\footnote{\url{https://pypi.org/project/nifty-ls/1.1.0/\#nifty_ls-1.1.0.tar.gz}}. The OGLE-IV photometry analyzed in Sect.~\ref{sec:comparison} is publicly available through the OGLE Collection of Variable Stars, with the RR~Lyrae stars at \url{https://ogledb.astrouw.edu.pl/~ogle/OCVS/rrlyr_query.php} and the eclipsing binaries at \url{https://ogledb.astrouw.edu.pl/~ogle/OCVS/ecl_query.php}.

\begin{acknowledgements}
This research was funded in part by the National Science Center, Poland, grant Sonata Bis 2021/42/E/ST9/00038 to Rados\l{}aw Poleski.\\
Author contributions:
KR: Conceptualization, Data Curation, Investigation, Methodology, Software,
Validation, Visualization, Formal analysis, Writing -- original draft, Writing
-- review \& editing.
RP: Conceptualization, Validation, Writing -- review \& editing, Supervision.\\
We thank Aleksander Schwarzenberg-Czerny for helpful comments that let us improve the initial version of the manuscript. We are also grateful to Sean League for finding issues with the initial version of the code.
\end{acknowledgements}

\bibliographystyle{aa}
\bibliography{aa}

\end{document}